\documentclass[conference]{IEEEtran}
\IEEEoverridecommandlockouts
\usepackage{dblfloatfix}
\usepackage{graphics} 
\usepackage{epsfig} 
\usepackage{mathptmx} 
\usepackage{times} 
\usepackage{amsmath} 
\usepackage{amssymb}  
\usepackage{multirow}
\usepackage{float,soul,xcolor}
\usepackage{cleveref}
\usepackage[flushleft]{threeparttable}
\usepackage{booktabs,comment}
\usepackage{cite}
\usepackage[font=footnotesize]{caption} 
\usepackage{subcaption}
\usepackage[symbol]{footmisc}

\def\BibTeX{{\rm B\kern-.05em{\sc i\kern-.025em b}\kern-.08em
    T\kern-.1667em\lower.7ex\hbox{E}\kern-.125emX}}

\begin{document}
\title{An Accurate and Hardware-Efficient Dual Spike Detector for Implantable Neural Interfaces}
\author{\IEEEauthorblockN{Xiaorang Guo\IEEEauthorrefmark{1}\IEEEauthorrefmark{2}\IEEEauthorrefmark{3},
MohammadAli Shaeri\IEEEauthorrefmark{1} and Mahsa Shoaran\IEEEauthorrefmark{1}}
\IEEEauthorblockA{xiaorang.guo@tum.de\quad\{mohammad.shaeri, mahsa.shoaran\}@epfl.ch}
$^\ast$Institute of Electrical and Micro Engineering, Center for Neuroprosthetics, EPFL, 1202 Geneva, Switzerland\\
$^\dagger$Faculty of Electrical and Computer Engineering, Technische Universität Dresden, 01069 Dresden, Germany\\
\IEEEauthorrefmark{3}Chair of Computer Architecture and Parallel Systems, Technische Universität München, 85748 Garching, Germany
}

\maketitle
\begin{abstract}
Spike detection plays a central role in neural data processing and brain-machine interfaces (BMIs). 
A challenge for future-generation implantable BMIs is to build a spike detector that features both low hardware cost and high performance.
In this work, we propose a novel hardware-efficient and high-performance spike detector for implantable BMIs.
The proposed design is based on a dual-detector architecture with adaptive threshold estimation. 
The dual-detector comprises two separate TEO-based detectors that distinguish a spike occurrence based on its discriminating features in both high and low noise scenarios.
We evaluated the proposed spike detection algorithm  on the Wave\_Clus dataset. It achieved an average detection accuracy of 98.9\%, 
and over 95\% in high-noise scenarios, ensuring the reliability of our  method.
When realized in hardware with a sampling rate of 16kHz and 7-bits resolution, the detection accuracy is 97.4\%.
Designed in 65nm TSMC process, a 256-channel detector based on this architecture occupies only 682$\rm \bf\mu m^2/Channel$ and consumes 0.07$\rm \bf \mu W/Channel$, improving over the state-of-the-art spike detectors by 39.7\% in power consumption and 78.8\% in area, while maintaining a high accuracy. 
\end{abstract}


\section{Introduction}
In a brain network, neurons mainly communicate through brief electrical pulses known as action potentials or spikes~\cite{Kandel2012}.
Traditionally, a brain-machine interface (BMI) records multi-channel (100--200) neural activity from brain regions associated with a target task. 
Next,  spike occurrences are detected as the preliminary processing step for neural decoding.
Future-generation BMIs, on the other hand, will be implantable high-density systems \cite{shin2022256} with the capability of on-chip data processing~\cite{Shaeri2020SFS, yoo2021neural, shaeri2022challenges, shin2022256}.
The detection accuracy should be high in various signal scenarios to be reliable for  clinical use. 
Furthermore, the design should be low-power and area-efficient to be integrable on next-generation highly-miniaturized neural microchips \cite{zhu2021closed}. 

To date, various spike detectors have been proposed for implantable BMIs \cite{yang2015adaptive, Nabar2009TEO, fiorelli2020charge, gagnon2016wireless, zhang2021adaptive, TEO_yuning, DWIVEDI201887, xu2019unsupervised, seong2021multi,dvt14}.
The simple \textit{absolute thresholding (AT)} method compares the signal amplitude with a predefined threshold level \cite{gagnon2016wireless}.
Despite being hardware-efficient, this method is highly sensitive to noise, 
which leads to poor performance at low signal-to-noise ratios (SNRs).
To address this issue, the \textit{dual vertex threshold (DVT)} algorithm  uses a positive and a negative threshold for spike detection 
\cite{dvt14,seong2021multi}.
Detectors based on data transformation techniques have also been proposed to improve  performance. 
In \cite{fiorelli2020charge}, the moving average energy (MAE), which calculates the instant signal energy over a sliding window,  was proposed for spike detection.
The \textit{Teager energy operator (TEO)} (\textit{a.k.a.}, nonlinear energy operator, NEO) and its variants \cite{zhang2021adaptive, TEO_yuning, DWIVEDI201887} are among the popular data transformation techniques widely used in spike detectors. 
Moreover, time-frequency analysis methods such as \textit{discrete wavelet transform (DWT)} and \textit{stationary wavelet transform (SWT)} have been reported for spike detection \cite{6897966, yang2015adaptive, Shaeri2022Xform}.
\
Other designs combined wavelets and TEO to achieve a high performance \cite{Nabar2009TEO,lieb2017stationary}.
However, such methods 
are computationally complex for integration on implantable devices with limited hardware resources \cite{zhu2020closed,shoaran2018energy, zhu2020resot}.

Here, we present the design and implementation of a hardware-efficient dual spike detector with online threshold estimation.
The dual detector smooths the neural signal to reduce its high-frequency noise content. 
Next,  TEO transformation is applied to the original and smoothed signals to enhance spike detectability. 
A spike is detected if either of the aforementioned signals exceeds the associated threshold level.
The proposed dual detector achieves a high detection accuracy even in high-noise scenarios thanks to the dual-path detection method. 
To show the scalability of the design, we implemented a 256-channel digital spike detector based on this architecture, achieving state-of-the-art accuracy and hardware efficiency.
The remainder of this paper is organized as follows.
We introduce the proposed detection algorithm and adaptive thresholding method in Section II.
Section III presents the hardware implementation.
Simulation results are presented in Section~IV, followed by a conclusion in Section V.

\begin{figure}[t]
\captionsetup{justification=raggedright,singlelinecheck=false}
    \vspace{-3mm}
    \centering
    \includegraphics[width=\linewidth]{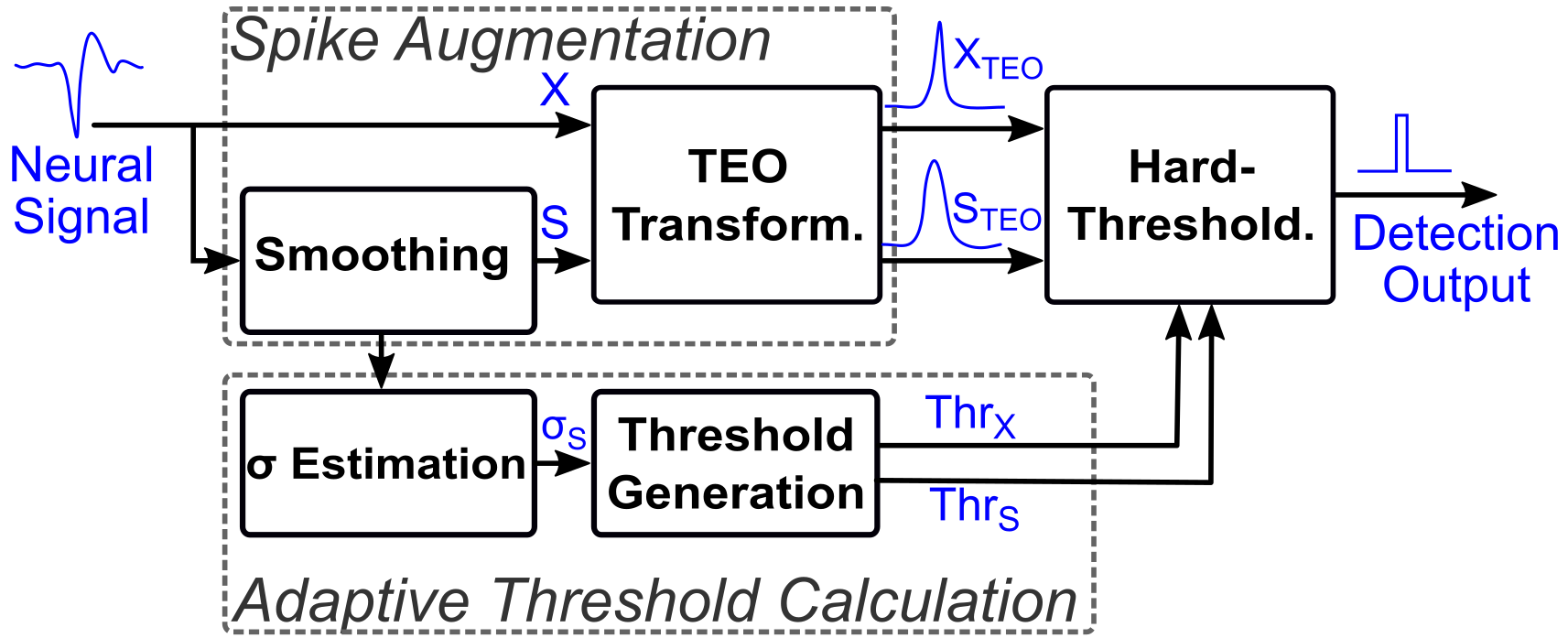}
    \caption{The generic block diagram of the proposed dual spike detector. 
  }
  \label{fig:whole_proc}
\vspace{-3mm}
\end{figure}

\section{Spike Detection Method}
The proposed dual-detection method comprises two separate spike detectors working in parallel. 
In one path,
the TEO of the input neural signal is calculated to highlight sharp peaks, followed by an adaptive thresholding procedure to distinguish the spikes from background noise.
The TEO transformation is formulated as follows:
\begin{flalign} \label{Equation: TEO}
&\mathcal{T}\{X\}[k] = X[k]^2 - X[k+1]X[k-1], 
\end{flalign}
where $X[k]$ and $\mathcal{T}\{X\}[k]$ represent the input and TEO signals, respectively \cite{Shaeri2022Xform}.
Since TEO accentuates sharp signal variations, it is only helpful when the high-frequency background noise is limited.
To detect spiking activity even in the presence of high-frequency noise, we included a second data path in our design.
In the second path, neural signals are first smoothed to reduce the high-frequency noise. 
The smoothed signal is then fed to a TEO-based spike detector.
The larger the window size, the better the smoothing function can be performed.
However, a large window size demands a larger memory  
that increases hardware utilization.
Considering the trade-off between performance and hardware cost, we set the window size to two samples in this design. 
Combining the results from both paths, a spike event will be raised if either detector captures above-threshold activities. 
Fig. \ref{fig:whole_proc} shows the block diagram of the proposed method.
As shown later in this paper, the dual-detection strategy leads to a significant improvement in detection accuracy, particularly in high-noise scenarios.

\begin{figure}[t]
	\centering
	\includegraphics[width=.4\textwidth]{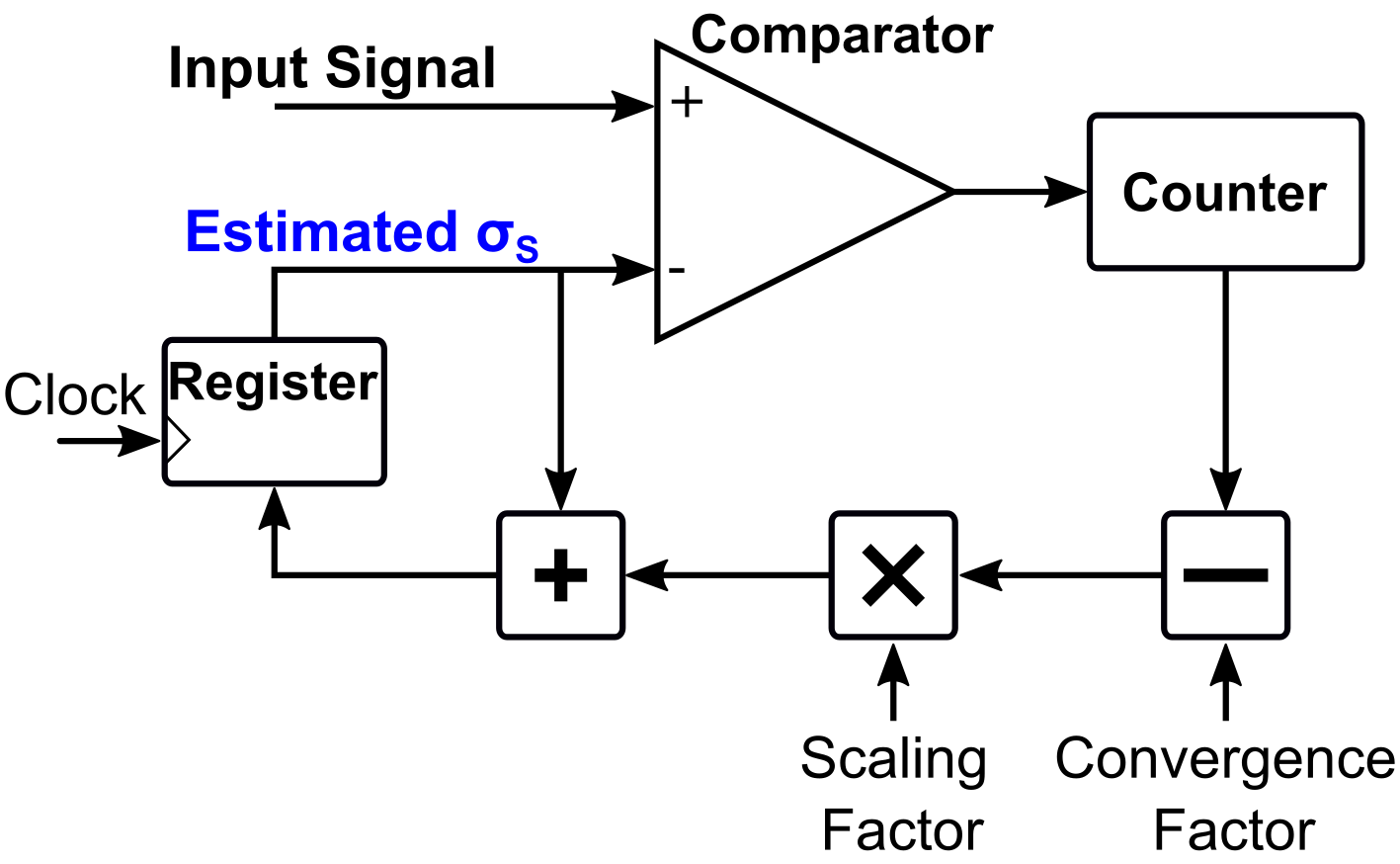}
	\caption{Block diagram of the on-chip standard deviation estimator for adaptive threshold calculation.
	}
	\label{fig:block_adpthr}
\vspace{-4mm}
\end{figure}

In a neural interface, the recorded signals are susceptible to drift over time.
In order to adjust the detection threshold to various changes in the signal, 
we designed an adaptive threshold estimator to calculate the threshold level in an online fashion on the chip. 
Due to the different characteristics of TEO-augmented signals ($X_{TEO}$, $S_{TEO}$), specific threshold levels are computed for each signal.
We formulate the threshold levels as follows: 
\begin{equation}
\begin{aligned}
& Thr_X = C_{1}\times \sigma_S,\\
& Thr_S = C_{2} \times \sigma_{S} + C_{3} \times \sigma_{S}^2, 
\end{aligned}
\label{eq:threshold}
\end{equation}
where $Thr_X$ and $Thr_S$ indicate the threshold levels for $X_{TEO}$ and $S_{TEO}$ signals (shown in Fig. \ref{fig:whole_proc}), respectively. $\sigma_S$ denotes the estimated standard deviation of signal `$S$', $C_1$, $C_2$ and, $C_3$ are the coefficients calculated via co-optimization of the hardware and detection accuracy.
The typical values of these coefficients are in the form of powers of 2 and can be implemented with only a few shifts and additions rather than complex multiplications.

To calculate the statistical 
`standard deviation' or std, many signal samples are required which could increase the hardware complexity.
The number of data samples higher than std in a large neural dataset is proportional to the statistical standard deviation \cite{yang2015adaptive}.
Therefore, based on this concept, we modified the estimation method introduced in \cite{yang2015adaptive} to improve the hardware efficiency for online std calculation. 
\
Fig. \ref{fig:block_adpthr} illustrates the block diagram of the std estimator.
The smoothed signal is first 
compared with an initial, predefined value of $\sigma_S$. 
The counter then calculates the number of samples higher than $\sigma_S$.
This one-bit stream is subsequently accumulated every 256 clock cycles, and the result is subtracted by a `convergence factor' that is determined based on the empirical data distribution. 
Following scaling, $\sigma_S$ will be updated by the difference between the number of $\sigma_S$-exceeded samples and the associated convergence factor through the feedback loop, as shown in Fig.~\ref{fig:block_adpthr}.
In other words, this feedback loop is designed to ensure the convergence of the number of samples exceeding $\sigma_S$ 
to the convergence factor.
The small values of `convergence factor' and `scaling factor' could make the $\sigma_S$-estimation process more stable, 
but at the cost of higher latency.
Here, we chose 20 and 0.001 
for the convergence and scaling factors, respectively.
Following convergence, we calculate two separate threshold values based on Eq. \ref{eq:threshold}.

\begin{figure}[t]
\captionsetup{justification=raggedright,singlelinecheck=false}
	\centering
    \includegraphics[width=.95\linewidth]{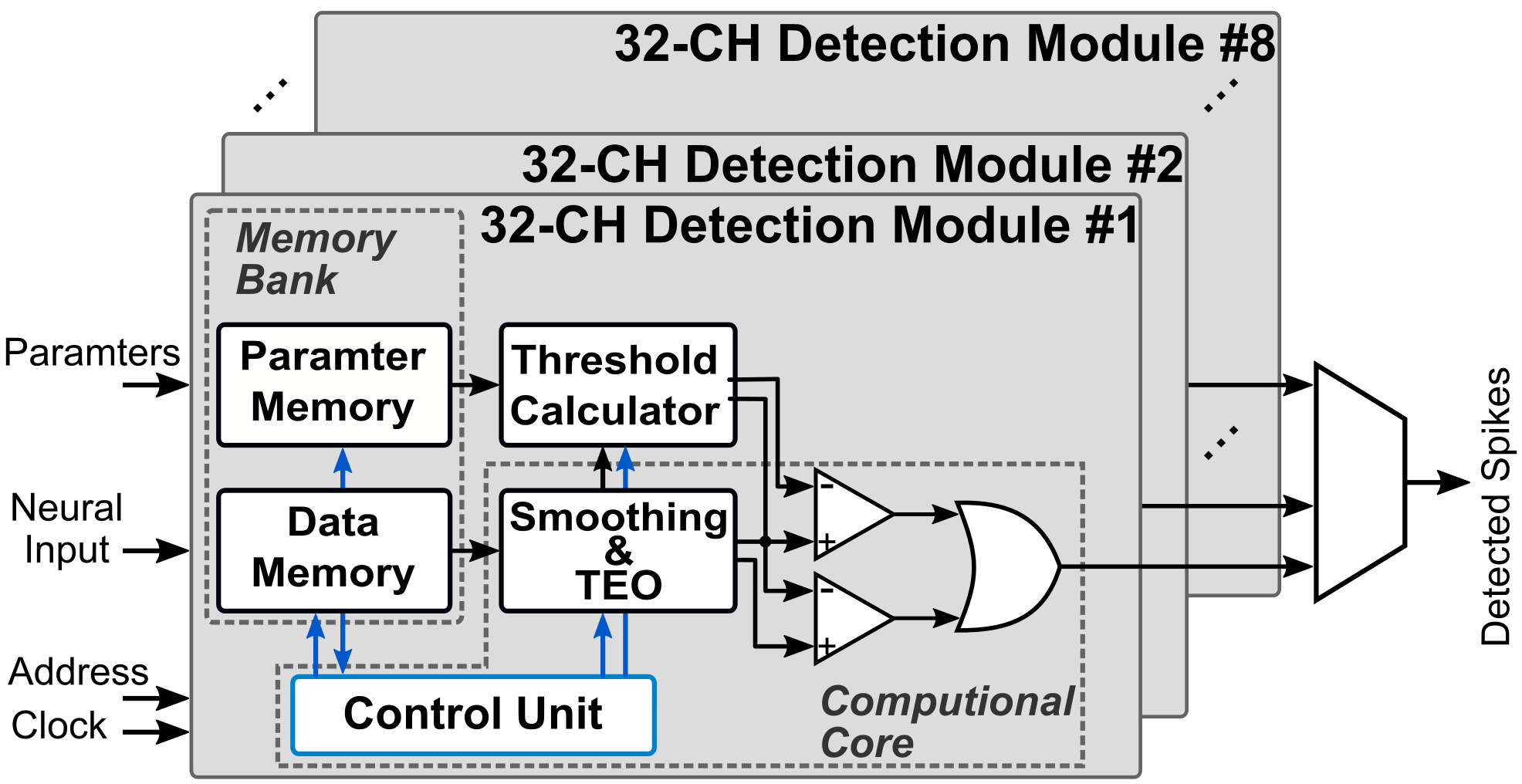}
	\caption{The hardware architecture of the 256-channel dual spike detector.
	}
	\label{fig:HardwareArchitecture}
\end{figure}
\begin{figure}[t!]
\captionsetup{justification=raggedright,singlelinecheck=false}
\vspace{-3mm}
\centering
	\includegraphics[width=0.9\linewidth]{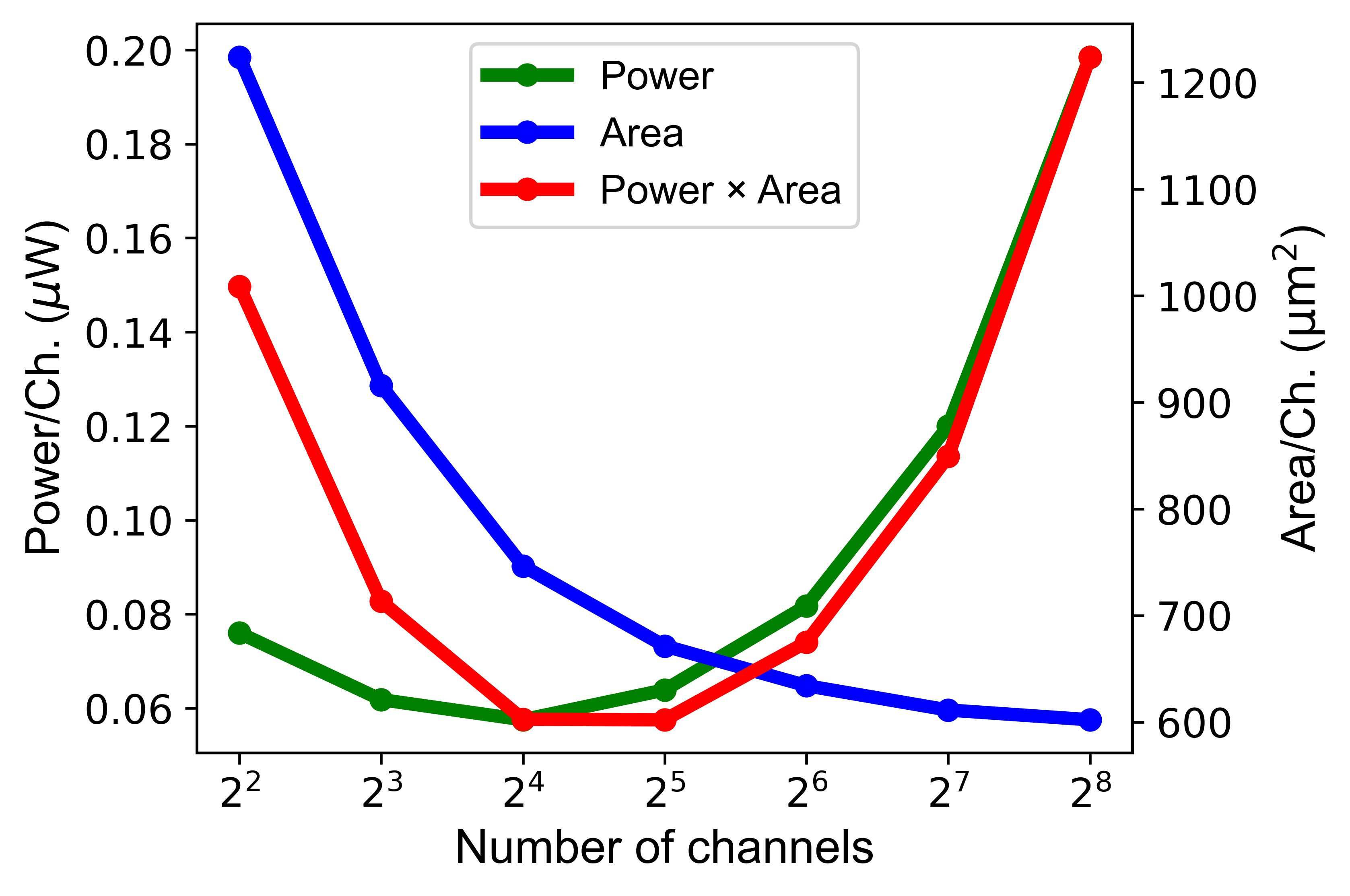}
	\caption{Hardware design optimization via area and power co-analysis. 
	}
\label{fig:area_power/ch}
\vspace{-6mm}
\end{figure}

\begin{figure*}[t]
\vspace{-8mm}
\centering
\subfloat[]{ \includegraphics[clip,width=.32\textwidth]{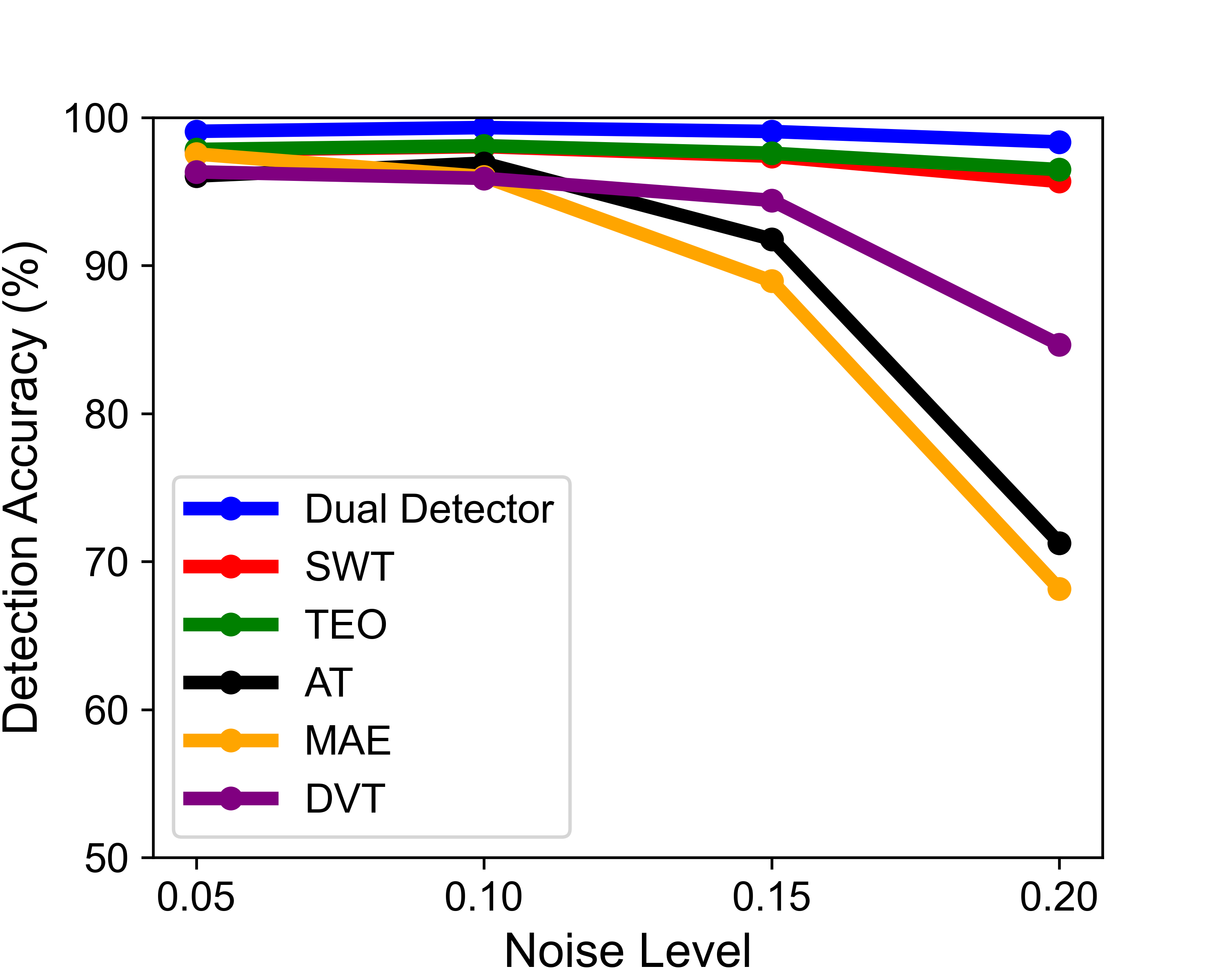} \vspace{-1.5mm}}
\subfloat[]{ \includegraphics[clip,width=.32\textwidth]{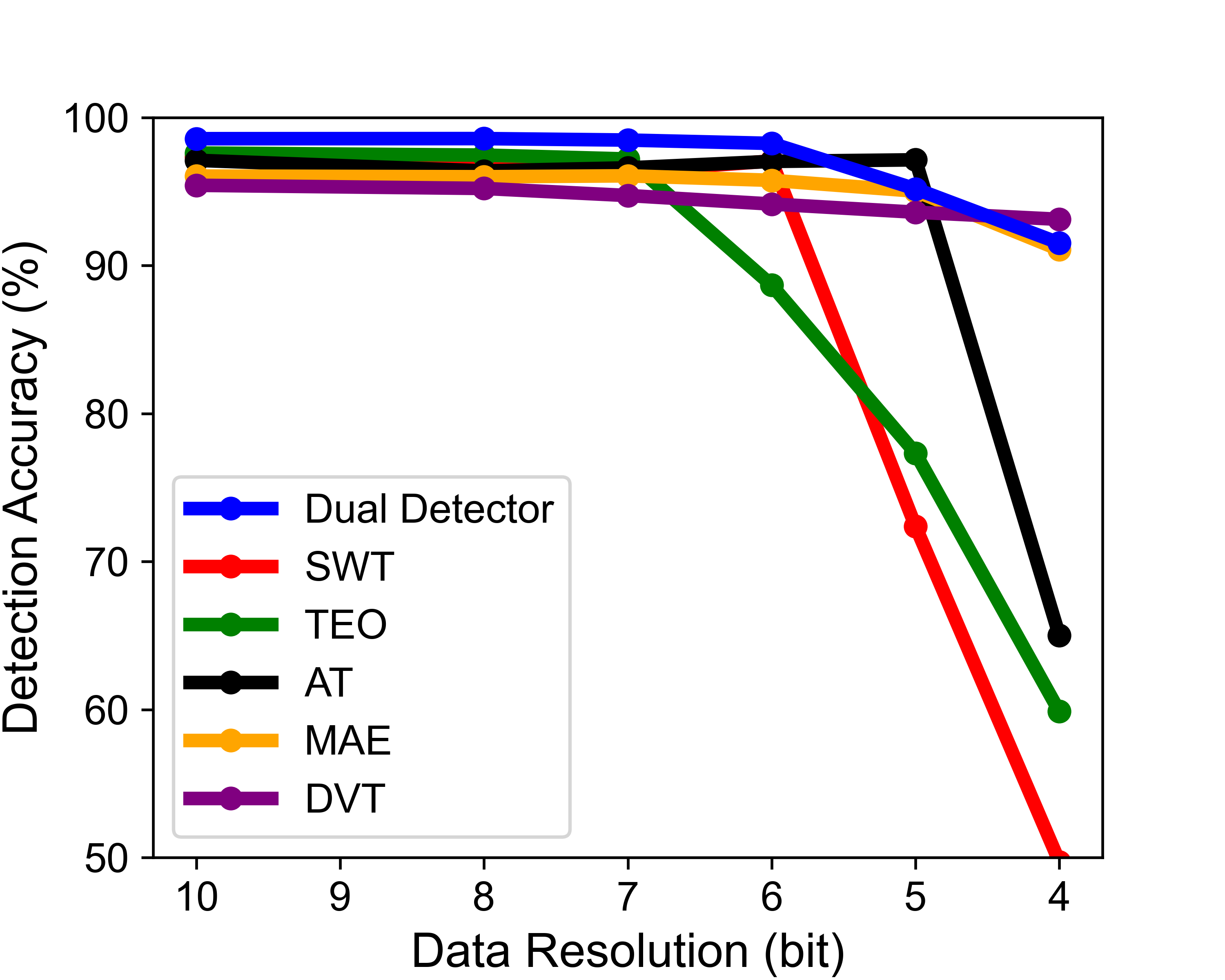} \vspace{-1.5mm}} 
\subfloat[]{ \includegraphics[clip,width=.32\textwidth]{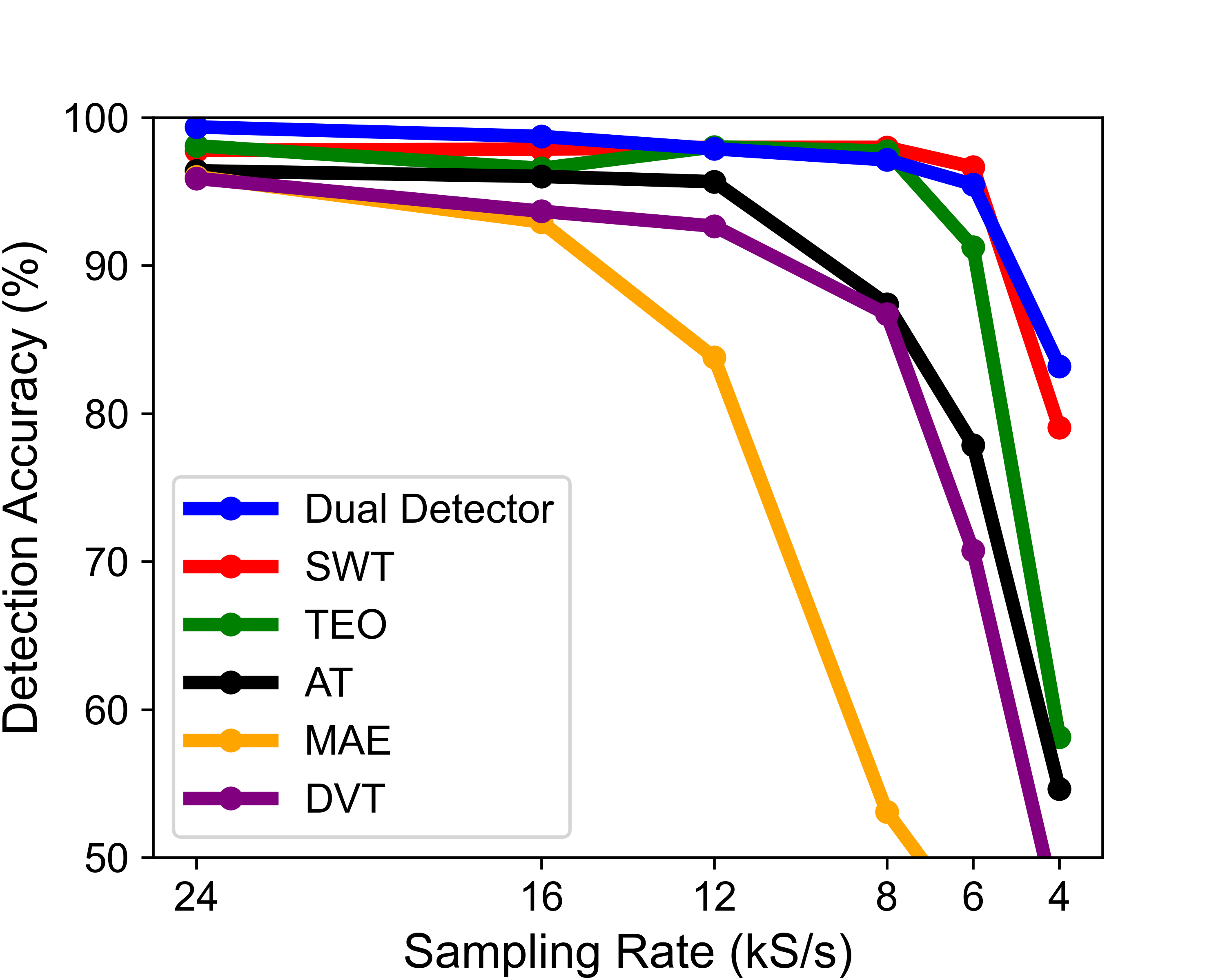} \vspace{-1.5mm}}
\vspace{-1.5mm}
\caption{The mean detection accuracy vs. (a) noise level, (b) data resolution, and (c) sampling rate for the proposed dual-detector and previous methods (algorithm-level simulation). For (b) and (c) the noise level is fixed at 0.1.
}
\label{fig:da_rs_sr}
\vspace{-2mm}
\end{figure*}

\section{Hardware Implementation}
Fig. \ref{fig:HardwareArchitecture} illustrates the modular architecture of the proposed 256-channel dual spike detector.
In this design, arithmetic operations are executed sequentially on the incoming channel data (i.e., 256 digitized channels). 
In order to reduce the hardware cost, the computational blocks such as TEO and threshold calculator are shared among every 32 channels. 
Here, five register banks store the channel data and threshold levels.
Also, a multiplexer is used to consolidate the outputs of the 32-channel modules.
As a result, the 1-bit data stream (detected spikes) will be generated via multiplexing the module outputs in the time domain. 
\
The original ($X$) and smoothed ($S$) neural signals are truncated to 7 and 6 bits, respectively, to further improve the hardware efficiency. 
Moreover, the TEO outputs are  truncated to 8-bit ($X_{TEO}$) and 9-bit ($S_{TEO}$) signals in order to reduce the memory requirements 
for thresholds calculation.
This can significantly improve the hardware efficiency at the cost of near $1\%$ drop in detection accuracy.

As shown in Fig. \ref{fig:area_power/ch}, there is a trade-off between the occupied area and power consumption of the proposed 256-channel spike detector.
Hardware sharing decreases the area per channel as the channel count increases. 
Similarly, the power per channel decreases as a function of channel count at low channel densities  (i.e., $<$16). This trend changes at higher channel counts where the dynamic power begins to dominate the system power.
By optimizing the area-power product, we can find  the optimal number of channels for hardware sharing (32). 
The 256-channel detector thus contains eight modules operating in parallel, each handling 32 channels. 
The system is clocked at 4MHz, while the 32-channel detectors are activated alternatively to reduce dynamic power.

\section{Results}
We used the Wave\_Clus dataset to assess the performance of our spike detector \cite{quiroga2004unsupervised}. 
The Wave\_Clus dataset includes four different subsets named Easy1, Easy2, Difficult1, and Difficult2 sampled at 24kS/s. 
Each subset has different noise levels, ranging from 0.05 to 0.2 (for Easy1, it can be up to 0.4).
The noise level is defined as the standard deviation of noise relative to the average magnitude of spikes.
To evaluate the spike detection performance, we used the ratio of correctly detected spikes (true positives, TP) to the total number of detected (TP+FP) and missed (FN)
spikes 
as  accuracy criteria. 

Fig. \ref{fig:da_rs_sr}(a) shows the detection accuracy of the proposed algorithm and previous methods reported in literature versus the noise level of the input signal.
When the noise level is at its minimum (0.05), all methods could detect the spikes with high accuracy ($>$95\%).
However, by increasing the noise level, the performances of MAE, AT, and DVT considerably decreased. 
The mean accuracies of MAE, AT, and DVT are 87.7\% (97.5-68.1\%), 89.0\% (96.1-71.2\%), and 92.8\% (96.3-84.7\%), respectively. 
TEO and SWT achieved a high mean accuracy of 97.5\% (97.8-96.5\%), and 97.2\% (97.8-95.7\%). 
The proposed dual detection method (with a mean accuracy of 98.9\%)  outperformed all other methods, with accuracies ranging from 99.1\% (at high SNR) to 98.4\% (at low SNR).  

Furthermore, we investigated the impact of input data resolution and sampling rate on the detection performance.
Fig. \ref{fig:da_rs_sr}(b) shows the detection accuracy versus resolution 
at a noise level of 0.1.
The performances of TEO, AT and SWT were high at high resolutions, with the performance rapidly dropping for lower resolutions. 
This analysis  shows that the proposed method, DVT and MAE are less sensitive to  this parameter, obtaining  accuracies above 90\% even at 4-bit resolution.
\
Fig.~\ref{fig:da_rs_sr}(c) further illustrates the detection accuracy versus  sampling rate for various spike detection methods.
As shown in this figure, our proposed method and SWT are less sensitive to   sampling rate compared to other methods.

%


\begin{table*}[t]
\vspace{-0mm}
\centering
\caption{
The performance summary of the proposed method and  state-of-the-art spike detectors.}
\label{table:percomparison}
\resizebox{1\textwidth}{!}{
\begin{threeparttable}
\begin{tabular}{|lp{0.16\textwidth}|p{0.08\textwidth}|p{0.08\textwidth}|p{0.08\textwidth}|p{0.08\textwidth}|p{0.08\textwidth}|p{0.08\textwidth}|p{0.08\textwidth}|p{0.08\textwidth}|p{0.08\textwidth}|p{0.08\textwidth}|}
\hline
\multicolumn{2}{|l|}{\textbf{Spike Detector}} &
  This work &
  \cite{gagnon2016wireless} &  \cite{yang2015adaptive} & \cite{fiorelli2020charge} & \cite{TEO_yuning} & \cite{DWIVEDI201887} & \cite{xu2019unsupervised} & \cite{seong2021multi} & \cite{Yoon2021} & \cite{9681264} \\ \hline
\multicolumn{2}{|l|}{\textbf{CMOS Process} (nm)} &   65 & 130 & 130 & 180 & 130 & 180 & 40 & 40 & 65 & 22 \\ \hline
\multicolumn{2}{|l|}{\textbf{Method}} & Dual-detection &   AT  &   SWT &   MAE &   TEO &   ED\footnotemark[4] &   PBOTM\footnotemark[1] &   DVT &   N/A &   TEO \\ \hline
\multicolumn{2}{|l|}{\textbf{Adaptive Threshold}} & \checkmark &   \checkmark &   \checkmark &   $\times$ &   \checkmark &   \checkmark &  $\times$  &   $\times$  &   N/A  &  $\times$ \\ \hline
\multicolumn{2}{|l|}{\textbf{Channel Count}} &   256 &   10 &   16 &   8 &   64(max) &   N/A &   16 &   16 &   1024 &   16 \\ \hline
\multicolumn{2}{|l|}{\textbf{Resolution (bits)}} &   7 &   16 &   6 &   8 &   10 &   N/A &   N/A &   12 &   10 &   9 \\ \hline
\multicolumn{2}{|l|}{\textbf{Sample Rate (kS/s)}} &   16 &   20  &   25 &   30 &   20 &   16 &   12 &   24 &   20 &   25 \\ \hline
\multicolumn{2}{|l|}{\textbf{Power per Channel ($\rm \mu W/Ch$)}} &  0.07 &  2.96 \footnotemark[2]  &  1.71 &  0.116 &  0.05\footnotemark[9] &  5.1 &  19.0\footnotemark[8] &  8.09\footnotemark[2] &  2.72 & 0.29\\ \hline
\multicolumn{2}{|l|}{\textbf{Area per Channel ($\rm mm^2/Ch$)}} &   $6.82\times 10^{-4}$ &  0.08\footnotemark[8]  & 0.014 &  0.27 &  $1.6\times 10^{-3}$\footnotemark[9] &  0.018 &  0.0175\footnotemark[8] &  N/A &  0.02 &  $3.22\times 10^{-3}$ \\ \hline
\multicolumn{2}{|l|}{\textbf{Accuracy}} &   97.4\% &   97.8\% \footnotemark[3]  &   98\%$\sim$99\%\footnotemark[5] &   97\%\footnotemark[3] &   95\%\footnotemark[3]&  95\% &   98.3\%\footnotemark[7]&   98.12\%\footnotemark[7] &   N/A &   N/A \\ \hline
\end{tabular}
\begin{tablenotes}
	\item[\textdaggerdbl] A different  dataset was used.
	\item[\textsection] ED represents the \textit{energy of derivative} method that is an approximated calculation of TEO designed in analog. 
	\item[\textasteriskcentered] PBOTM refers to the \textit{preselection Bayes optimal template matching} that simultaneously performs spike detection and spike sorting. 
	\item[\textparagraph] Overlapping spikes were excluded in the calculation of  detection accuracy.
    \item[\textdaggerdbl\textdaggerdbl] Memory blocks were excluded in the estimation of power and area. 
    \item[\textasteriskcentered\textasteriskcentered] False positives were not considered in the calculation of detection accuracy. 
    \item[\textdagger] Power is estimated based on the power breakdown in the paper.
    \item[\textdagger\textdagger] Power or area was reported for the whole design, which also includes the spike sorting block, compression block or others.
	\end{tablenotes} 
\end{threeparttable}} 
\vspace{-4mm}
\end{table*}


\begin{figure}[b]
\vspace{-8mm}
	\centering
    \includegraphics[width=.6\linewidth]{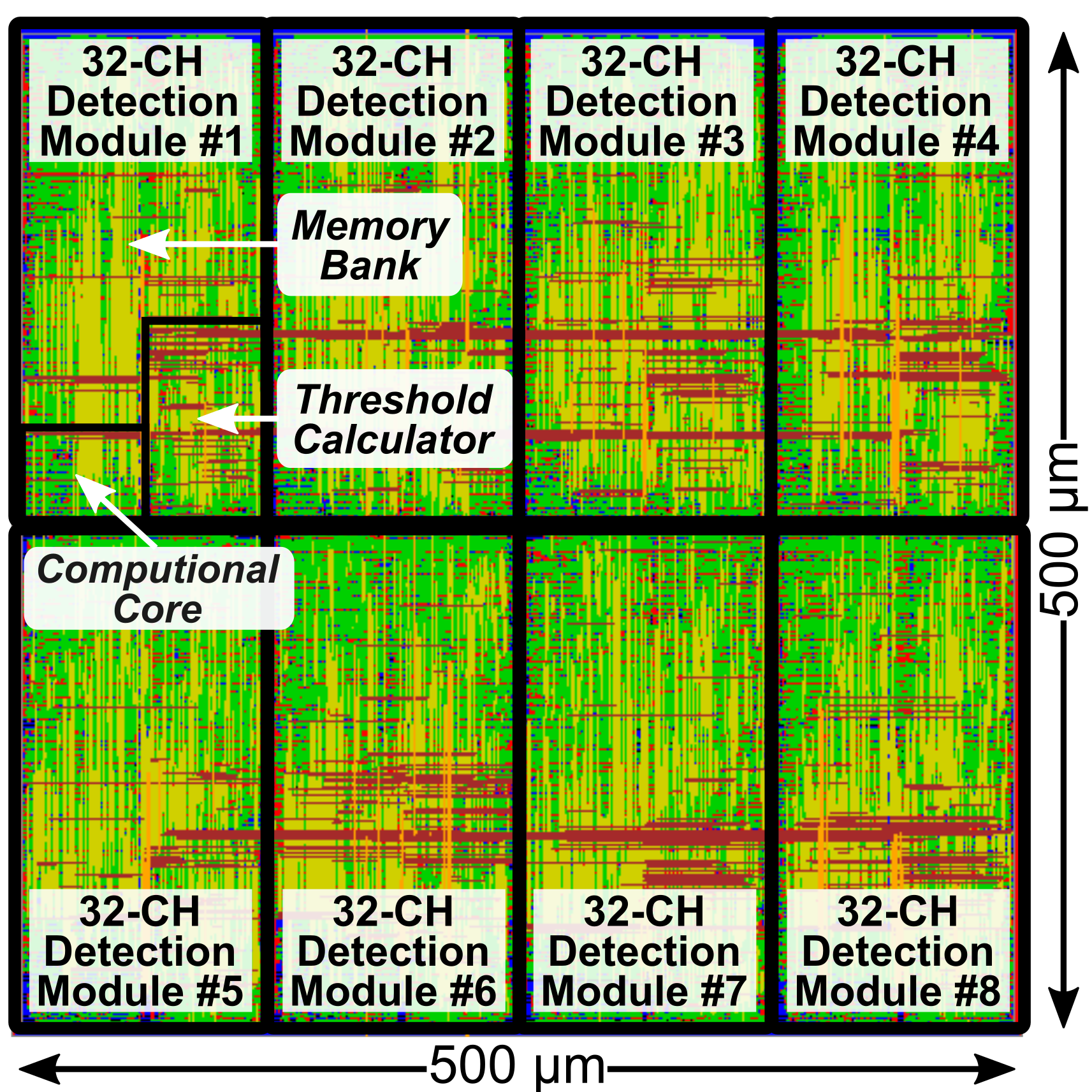}
	\caption{The physical layout of the 256-channel dual spike detector. The area of each 32-channel module is about 0.031$\rm mm^2$.
	}
	\label{fig:Layout}
	\vspace{-2mm}
\end{figure}

\begin{figure}[t!]
\vspace{-1mm}
\centering
\begin{subfigure}{.5\linewidth}
  \centering
  \includegraphics[width=1\linewidth]{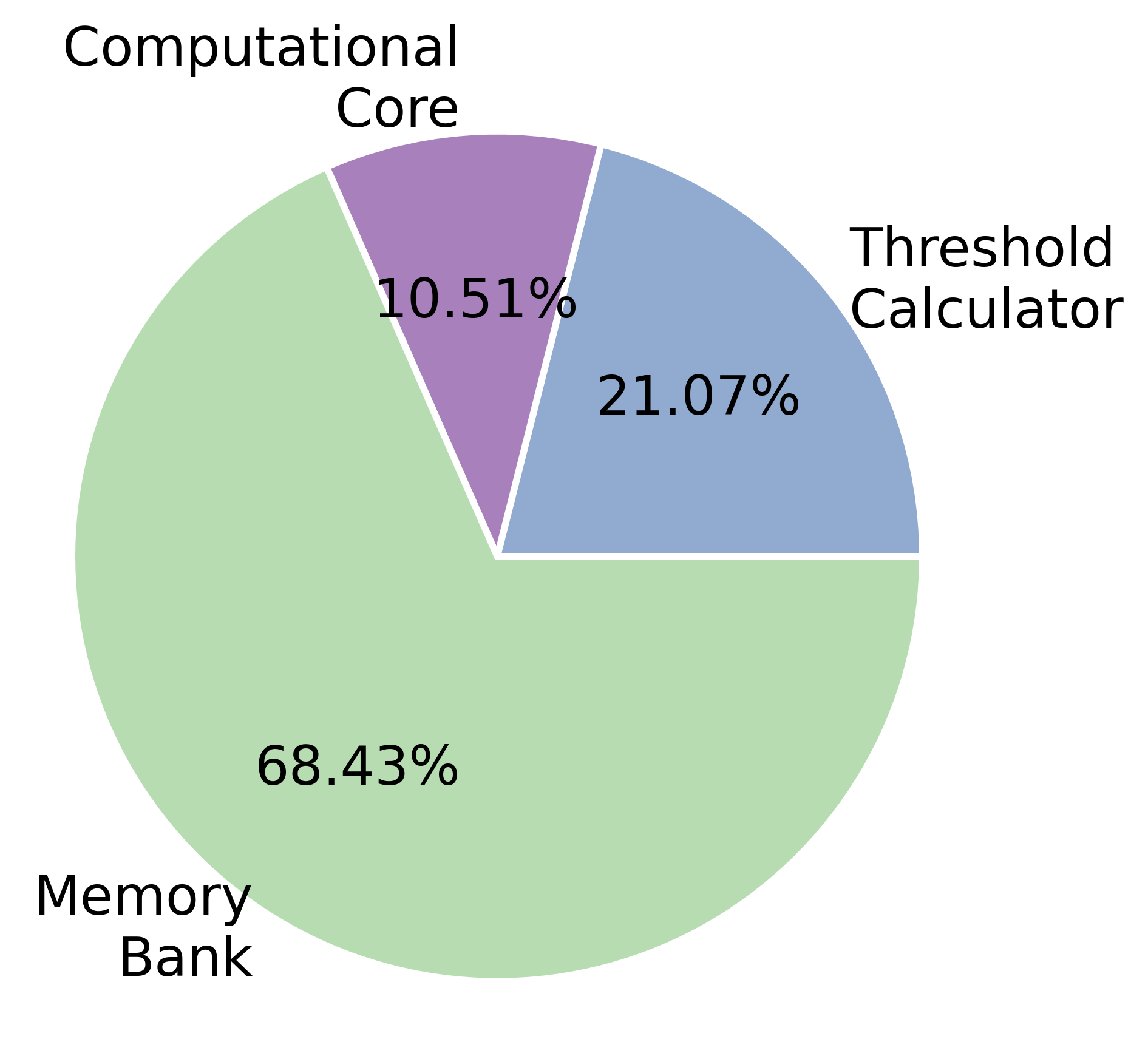}
  \caption{}
\end{subfigure}%
\begin{subfigure}{.5\linewidth}
  \centering
  \includegraphics[width=1\linewidth]{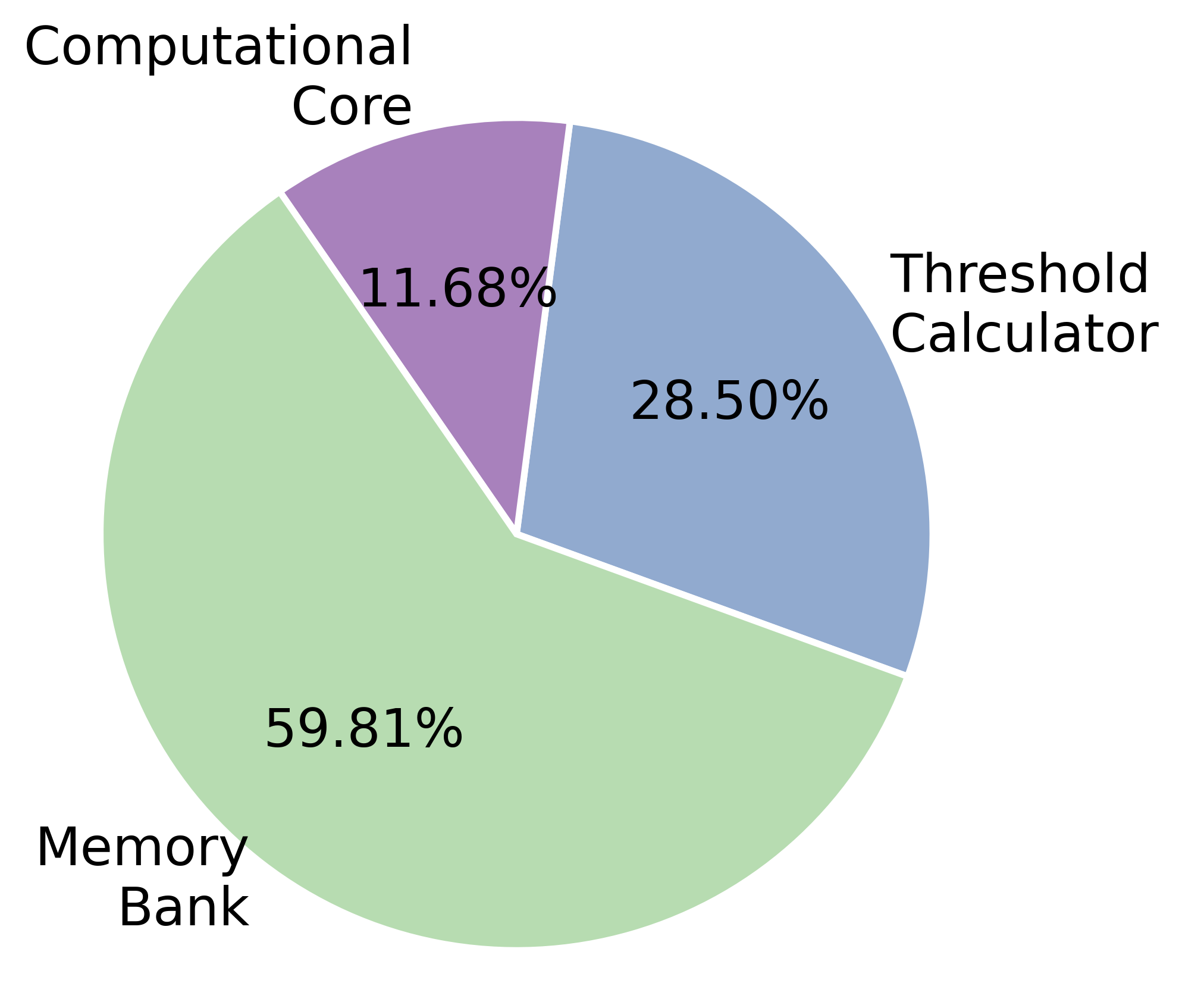}
  \caption{}
\end{subfigure}
\caption{(a) The area, and (b) power breakdowns of the proposed dual  detector. The memory bank contains 
register banks for storing the internal values, clock gating, and threshold estimation. 
The computational units represent the hardware responsible for  computations such as TEO, smoothing, and comparators. 
The threshold calculator contains the  logic to implement the architecture  in Fig. \ref{fig:block_adpthr} (excluding the registers).
}
\label{fig:breakdown}
\vspace{-5mm}
\end{figure}

Table. \ref{table:percomparison} summarizes the design specifications, detection accuracy, and  hardware-level performance of the proposed method and  state-of-the-art spike detectors. References \cite{gagnon2016wireless,fiorelli2020charge,Yoon2021,9681264} present  measured  results, while the others report  simulated hardware results.
We designed the 256-channel dual spike detector in a 65$\rm nm$ TSMC process, with an active area of 0.25$\rm mm^2$ and power consumption of 17.55$\rm \mu W$ at a 1.1V supply.
Fig. \ref{fig:Layout} shows the layout of the 256-channel dual spike detector introduced in Fig. \ref{fig:HardwareArchitecture}.
The simulated dynamic and leakage powers are 10.8$\rm \mu W$ and 6.74$\rm \mu W$ at 16kHz, respectively.
The area and power breakdowns  are shown in Fig.~\ref{fig:breakdown}.
The register banks (i.e., memory) are the dominant block in terms of both power and area usage,  consuming over 68\% of the hardware resources in this design. 

For a fair comparison against the state-of-the-art, the per-channel area and power consumption  of the proposed detector and other similar designs are  presented in Table \ref{table:percomparison}.
The area and power of the dual-detector are 682$\rm \mu m^2/Channel$ and 70$\rm nW/Channel$, respectively.
These results show 78.8\% and 39.7\% improvements in area utilization and power consumption, respectively, over the state-of-the-art spike detectors \cite{fiorelli2020charge,9681264}.
This comparison confirms that our proposed spike detector is the most hardware-efficient design reported so far.
This is a result of extensive hardware optimizations (e.g., block sharing and data truncation). 
It is worth mentioning that the power consumption of 50$\rm nW/Channel$ reported in \cite{TEO_yuning} did not include the power consumed by memory banks, which could be significant in high-density BMIs. 

From the standpoint of detection performance, the proposed dual detector achieved an overall accuracy of 97.4\% after hardware optimization. 
This performance is comparable to the state-of-the-art spike detectors while being achieved at a significantly lower hardware cost.
The high accuracy of $\sim$99\% reported in \cite{yang2015adaptive} was achieved by excluding the overlapping spikes in the dataset.
Moreover, the detection accuracies of \cite{xu2019unsupervised} and \cite{seong2021multi}, which are slightly higher than our detector, were calculated with a different definition (i.e., the ratio of true positives to the sum of true positives and false negatives). 
Thus, false positives were discarded in their calculations.
In addition, these performances were achieved at significantly higher hardware costs compared to our proposed detector.

\section{Conclusion}
This paper presents a novel spike detector that benefits from a dual detection architecture.
This algorithm achieved nearly 99\% accuracy in software simulations, and 97.4\%  using an efficient hardware implementation.
To improve the hardware efficiency, we optimized our system for low sampling rate and data resolution as well as  optimal channel count for hardware sharing. 
In  65nm TSMC process, the dual-detector occupies only 682$\rm \mu m^2/Channel$ and consumes only 0.07$\rm \mu W/Channel$,
making it a proper candidate for implantable BMIs.
The proposed spike detector outperforms current state-of-the-art spike detectors in terms of hardware efficiency. 


\footnotesize
\bibliographystyle{IEEEtran}
\bibliography{main.bbl}

\end{document}